\documentclass[10pt,pdftex]{iopart}
\usepackage{iopams}

\usepackage{tensind,graphicx,gensymb,bbm,wasysym} 
\usepackage{mathptmx}      
\usepackage{color}

\newcommand{\GV}{\textit{GeodesicViewer}}
\newenvironment{gvExercise}{\vspace*{0.1cm} \begin{minipage}[t]{4.5in}\textbf{Exercise:}\it}{\end{minipage}\vspace*{0.1cm}}
\newenvironment{gvResult}{\vspace*{0.1cm} \begin{minipage}[t]{4.5in}\textbf{Result:}}{\end{minipage}\vspace*{0.1cm}}
\newenvironment{gvConf}{\vspace*{0.1cm} \begin{minipage}[t]{4.5in}\textbf{Configure file: }\color{blue}}{\color{black}\end{minipage}\vspace*{0.1cm}}

\begin{document}
\tensordelimiter{?}

\title{Studying null and time-like geodesics in the classroom}

\author{Thomas M{\"u}ller}
\address{
  Visualisierungsinstitut der Universit{\"a}t Stuttgart (VISUS),\\
  Allmandring 19, 70569 Stuttgart, Germany
}
\ead{Thomas.Mueller@vis.uni-stuttgart.de}

\author{J{\"o}rg Frauendiener\footnote{and: Centre of Mathematics for Applications, University of Oslo, P.O. Box 1053, Blindern, NO-0316 Oslo, Norway}}
\address{
  Department of Mathematics \& Statistics, University of Otago,\\
  P.O. Box 56, Dunedin 9010, New Zealand
}
\ead{joergf@maths.otago.ac.nz}

\begin{abstract}
 In a first course of general relativity it is usually quite difficult for students to grasp the concept of a geodesic. It is supposed to be straight (auto-parallel) and yet it `looks' curved. In these situations it is very useful to have some explicit examples available which show the different behaviour of geodesics. In this paper we present the \GV, an interactive tool for studying the behaviour of geodesics in many different space-times. The geodesics can be represented in several ways, depending on the space-time in question. The use of a local reference frame and `Cartesian-like' coordinates helps the students to develop some intuition in various situations. We present the various features of the \GV\ in the form of readily formulated exercises for the students.
\end{abstract}


\submitto{\EJP}

\section{Introduction}
The intrinsic curvature of a space-time in general relativity is a concept that contradicts our every day experience of space and time. The necessary mathematical set of tools is difficult to learn and fairly abstract. To get some impression what a curved space-time means, is to study the behaviour of light rays and particles in free motion. In the geometric optics limit and for particles whose mass has no back-reaction on the curvature of space-time, light rays and particles in free motion can be represented by null and time-like geodesics, respectively.

For a first glimpse on how null and time-like geodesics behave, it may be sufficient to use off-the-shelf/standard software like for example Maple, Mathematica, or Octave. All of them could integrate the geodesic equation and show the geodesic as 2d- or 3d-plot. However, to explore the behaviour of geodesics, an interactive tool is indispensable.

\begin{figure}
 \centering
 \includegraphics[scale=0.25]{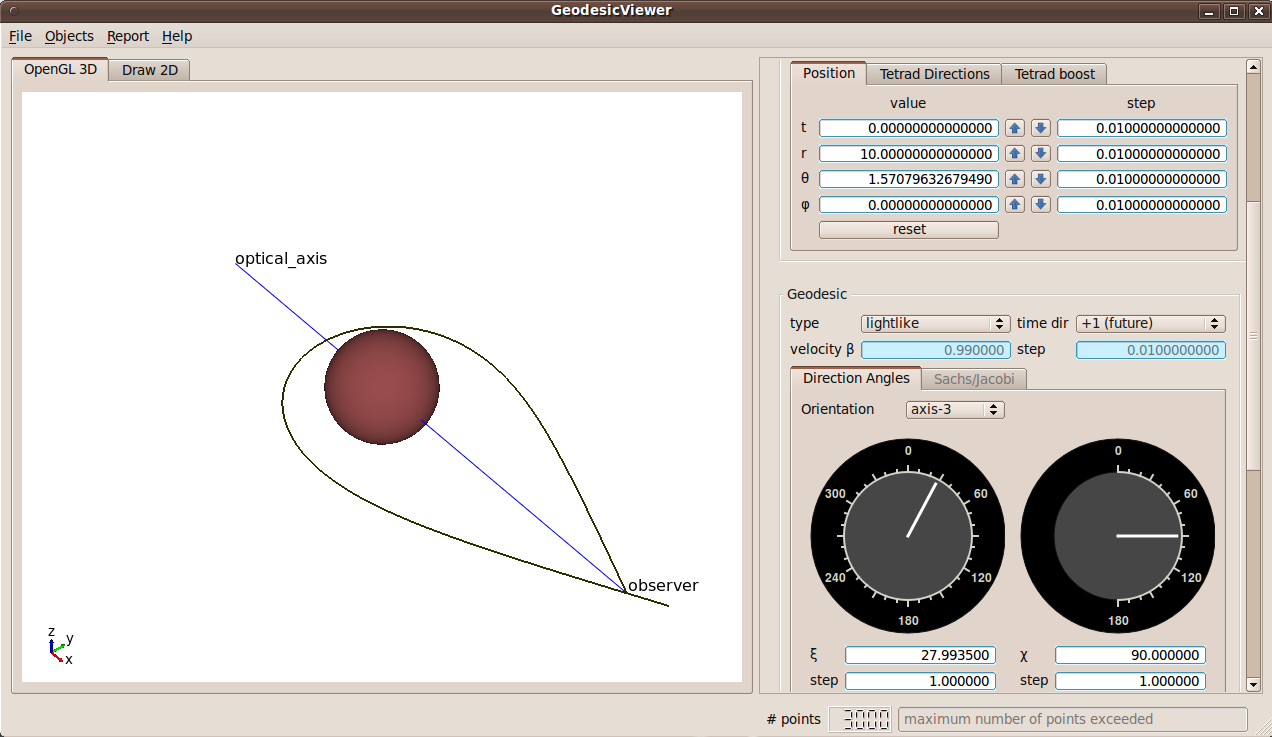}
 \caption{\label{fig:screenshot}Screenshot of the \GV{}'s user interface.}
\end{figure}

In this article we present the \GV~\cite{mueller2010a}, an interactive tool to thoroughly examine the behaviour of light-like and time-like geodesics in a space-time whose metric is provided analytically. The database of metrics is taken from the Motion4D library~\cite{mueller2009b}. The metrics with the corresponding Christoffel symbols and local tetrads are detailed in Ref.~\cite{mueller2009a}. The graphical user interface, see Fig.~\ref{fig:screenshot}, is written using the object-oriented, cross-platform application framework Qt\cite{qt}. The graphical 2D and 3D output is realized by means of the Open Graphics Library (OpenGL)\cite{opengl}. For the numerical integration of the geodesics, we use a standard fourth-order Runge-Kutta method and the integrators of the GNU Scientific Library (GSL)\cite{gsl}.

The structure of the paper is as follows. In Sec.~\ref{sec:GV} we give a short description of the \GV. Sec.~\ref{sec:schwarzschild} discusses the standard situations in the Schwarzschild space-time. Periodic orbits of time-like geodesics in black hole space-times are worked out in Sec.~\ref{sec:kerrPeriodic}. Secs.~\ref{sec:wormhole} and \ref{sec:multiBH} deal with the more exotic space-times like the Morris-Thorne wormhole and the extreme Reissner-Nordstr{\o}m diblack hole.

The \GV{} is freely available for Linux and Windows. The source code and several examples can be downloaded from \textit{www.vis.uni-stuttgart.de/relativity}.

\section{GeodesicViewer}\label{sec:GV}
The two main outputs of the \GV{} are 3D and 2D representations of the geodesic data. In the standard 3D representation,  a geodesic is depicted by means of pseudo-Cartesian coordinates, where the inherent coordinates are transformed into Cartesian coordinates as usual. If an embedding diagram is defined for a specific hypersurface of the space-time, the geodesic can also be represented in this form. (In principle, the user can implement any representation he wants.) The standard 2D representation of a geodesic is also given in pseudo-Cartesian coordinates, where now only a specific hypersurface is used. Another representation shows coordinate or velocity relations. For example, the radial coordinate could be plotted against the affine parameter. A third representation follows from the Euler-Lagrangian formalism, where an effective potential can be defined. This representation is particularly helpful to find bound orbits.

The numerous features of the \GV{} are described by means of examples which are formulated as exercises. The mathematical explanations are directed to the teacher. The exercises can be made either from scratch or a configure file can be prepared in advance, so that the student only has to change a few parameters. Each result description is accompanied by a configure file which holds the final result for reproduction in the \GV{}. To familiarize oneself with the graphical user interface of the \GV{}, there are several online tutorials. These can be worked through either alone or with guidance from the teacher.

\section{Basic examples in the Schwarzschild space-time}\label{sec:schwarzschild}
The prime example of general relativity is the Schwarzschild metric which we will give here in isotropic coordinates $x^{\mu}=(t,x,y,z)$. The line element $ds^2=g_{\mu\nu}dx^{\mu}dx^{\nu}$ reads
\begin{equation}
  ds^2 = -\left(\frac{1-\rho_s/\rho}{1+\rho_s/\rho}\right)^2c^2dt^2 + \left(1+\frac{\rho_s}{\rho}\right)^4\left[dx^2+dy^2+dz^2\right],
  \label{eq:schwMetricIso}
\end{equation}
where $\rho^2=x^2+y^2+z^2$, $\rho_s=GM/(2c^2)$ is the Schwarzschild radius, $G$ is Newton's constant, $M$ is the mass of the black hole, and $c$ is the speed of light. The transformation between the usual Schwarzschild radial coordinate $r$ and the isotropic radial coordinate $\rho$ is given by $r=\rho\left(1+\rho_s/\rho\right)^2$. If $M=0$, Eq.~(\ref{eq:schwMetricIso}) simplifies to the Minkowski metric in Cartesian coordinates.

Because of the spherical symmetry of the Schwarzschild space-time, we can restrict geodesics to the $xy$-plane. Then, the geodesic equations read
\begin{eqnarray}
 0 &= \ddot{t} + 2\Gamma_{tx}^t\dot{t}\dot{x} + 2\Gamma_{ty}^t\dot{t}\dot{y},\\
 0 &= \ddot{x} + \Gamma_{tt}^x\dot{t}^2 +\Gamma_{xx}^x\dot{x}^2 + 2\Gamma_{xy}^x\dot{x}\dot{y}+\Gamma_{yy}^x\dot{y}^2,\\
 0 &= \ddot{y} + \Gamma_{tt}^y\dot{t}^2 +\Gamma_{xx}^y\dot{x}^2 + 2\Gamma_{xy}^y\dot{x}\dot{y}+\Gamma_{yy}^y\dot{y}^2,
\end{eqnarray}
with the Christoffel symbols
\begin{eqnarray}
  \Gamma_{tt}^x = \frac{2c^2\rho^3\rho_s\left(\rho-\rho_s\right)x}{\left(\rho+\rho_s\right)^7}, \quad \Gamma_{tt}^y = \frac{2c^2\rho^3\rho_s\left(\rho-\rho_s\right)y}{\left(\rho+\rho_s\right)^7},\\
  \Gamma_{tx}^t = \frac{2\rho_s x}{\rho^3\left[1-\rho_s^2/\rho^2\right]}, \quad \Gamma_{ty}^t = \frac{2\rho_s y}{\rho^3\left[1-\rho_s^2/\rho^2\right]},\\
  \Gamma_{xx}^x = \Gamma_{xy}^y = -\Gamma_{yy}^x = -\frac{2\rho_s}{\rho^3}\frac{x}{1+\rho_s/\rho},\\
  \Gamma_{yy}^y = -\Gamma_{xx}^y = \Gamma_{xy}^x = -\frac{2\rho_s}{\rho^3}\frac{y}{1+\rho_s/\rho}.
\end{eqnarray}
Here, a dot represents the derivative with respect to the affine parameter $\lambda$, hence $\dot{t}=dt/d\lambda$.

\begin{figure}[htb]
 \centering
 \includegraphics[scale=1.0]{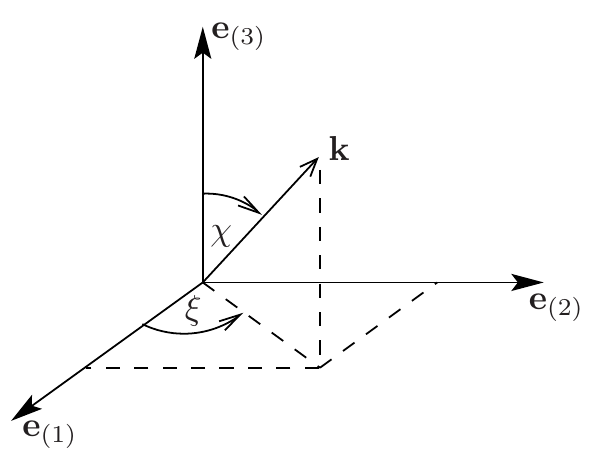}
 \caption{Initial direction $\mathbf{k}=k^{(i)}\mathbf{e}_{(i)}$ with respect to the local reference frame of an observer.}
 \label{fig:initDir}
\end{figure}

To integrate the geodesic equations, we need not only an initial position but also an initial direction. For this purpose, we first introduce the local reference frame $\{\mathbf{e}_{(i)}\}_{i=t,x,y,z}$ of an observer which defines a local Minkowskian system. The four base vectors $\mathbf{e}_{(i)}=\mathbf{e}_{(i)}^{\mu}\partial_{\mu}$ have to fulfill the orthonormality condition $g_{\mu\nu}\mathbf{e}_{(i)}^{\mu}\mathbf{e}_{(j)}^{\nu}=\eta_{(i)(j)}$ with $\eta_{(i)(j)}=\mbox{diag}(-1,1,1,1)$. Here, the most convenient choice for the local reference frame is the one which is adapted to the coordinates and the symmetries of the metric,
 \begin{eqnarray}
 \mathbf{e}_{(t)} &= \frac{1+\rho_s/\rho}{1-\rho_s/\rho}\frac{\partial_t}{c}, \qquad& \mathbf{e}_{(x)} = \left(1+\frac{\rho_s}{\rho}\right)^{-2}\partial_x,\\
  \mathbf{e}_{(y)} & = \left(1+\frac{\rho_s}{\rho}\right)^{-2}\partial_y, \qquad& \mathbf{e}_{(z)} = \left(1+\frac{\rho_s}{\rho}\right)^{-2}\partial_z.
 \end{eqnarray}
Now, an initial direction, $\mathbf{k}=k^{(i)}\mathbf{e}_{(i)}^{\mu}\partial_{\mu}=k^{\mu}\partial_{\mu}$, of a null or a time-like geodesic can be defined with respect to this local reference frame as shown in Fig.~\ref{fig:initDir}.

For a null geodesic we have $\mathbf{k}=\mathbf{e}_{(0)}+\sin\chi\cos\xi\mathbf{e}_{(1)}+\sin\chi\sin\xi\mathbf{e}_{(2)}+\cos\chi\mathbf{e}_{(3)}$.  The initial direction of a time-like geodesic, on the other hand, equals the initial four-velocity $\mathbf{k}=\mathbf{u}=\gamma c\mathbf{e}_{(0)}+\gamma\beta c\left(\sin\chi\cos\xi\mathbf{e}_{(1)}+\sin\chi\sin\xi\mathbf{e}_{(2)}+\cos\chi\mathbf{e}_{(3)}\right)$, where $\beta=v/c$ is the initial velocity $v$ scaled by the speed of light, and $\gamma=1/\sqrt{1-\beta^2}$. Because we restrict to geodesics in the $xy$-plane, we must set $\chi=90\degree$. The geodesic equations can now be solved using the initial position $x^{\mu}\big|_{\lambda=0}$ and the initial direction $\dot{x}^{\mu}\big|_{\lambda=0}=k^{\mu}$.

The numerical integration of the geodesics are accomplished by either a standard fourth-order Runge-Kutta method or by the integrators of the GNU Scientific Library. After each integration step, the geodesic is tested if it still fulfills the constraint equation $g_{\mu\nu}\dot{x}^{\mu}\dot{x}^{\nu}=\kappa c^2$ with $\kappa=0$ for light-like and $\kappa=-1$ for time-like geodesics. If this constraint is not fulfilled within a certain accuracy, the integration stops.

For the following examples, we set $G=c=M=1$. In physical units, this means that $G/c^2=1~ls/\mathcal{M}$ which equals one light second per mass unit $\mathcal{M}\approx 2.03\times 10^5M_{\astrosun}$. Hence, $M=1$ represents an object of $2.03\times 10^5$ solar masses and distances are measured in light seconds.

\subsection{Deflection of light}\label{subsec:deflection}
In the neighbourhood of a black hole, light rays are no longer straight lines but they are deflected due to the curvature of space-time. The size of the deflection depends on how close the light ray passes the black hole. 

\begin{gvExercise}
 Start the \GV{} and select the Schwarzschild metric in isotropic coordinates (SchwarzschildIsotropic) in the ``Metric/Integrator/Cons\-tants'' window. Set the position of an observer in the ``Local Tetrad'' window to $(t=0,x=6,y=6,z=0)$. With respect to his local reference frame, the observer starts light rays in the negative $y$-direction, $\mathbf{k}=\mathbf{e}_{(0)}-\mathbf{e}_{(2)}$. For that, set the direction angle $\xi=180\degree$ within the ``Geodesic'' window. Modify the initial position $y$ and describe the behaviour of the light rays. What is the closest approach where the light ray can still escape from the black hole?
\end{gvExercise}

\begin{figure}[htb]
 \centering
 \includegraphics[scale=0.5]{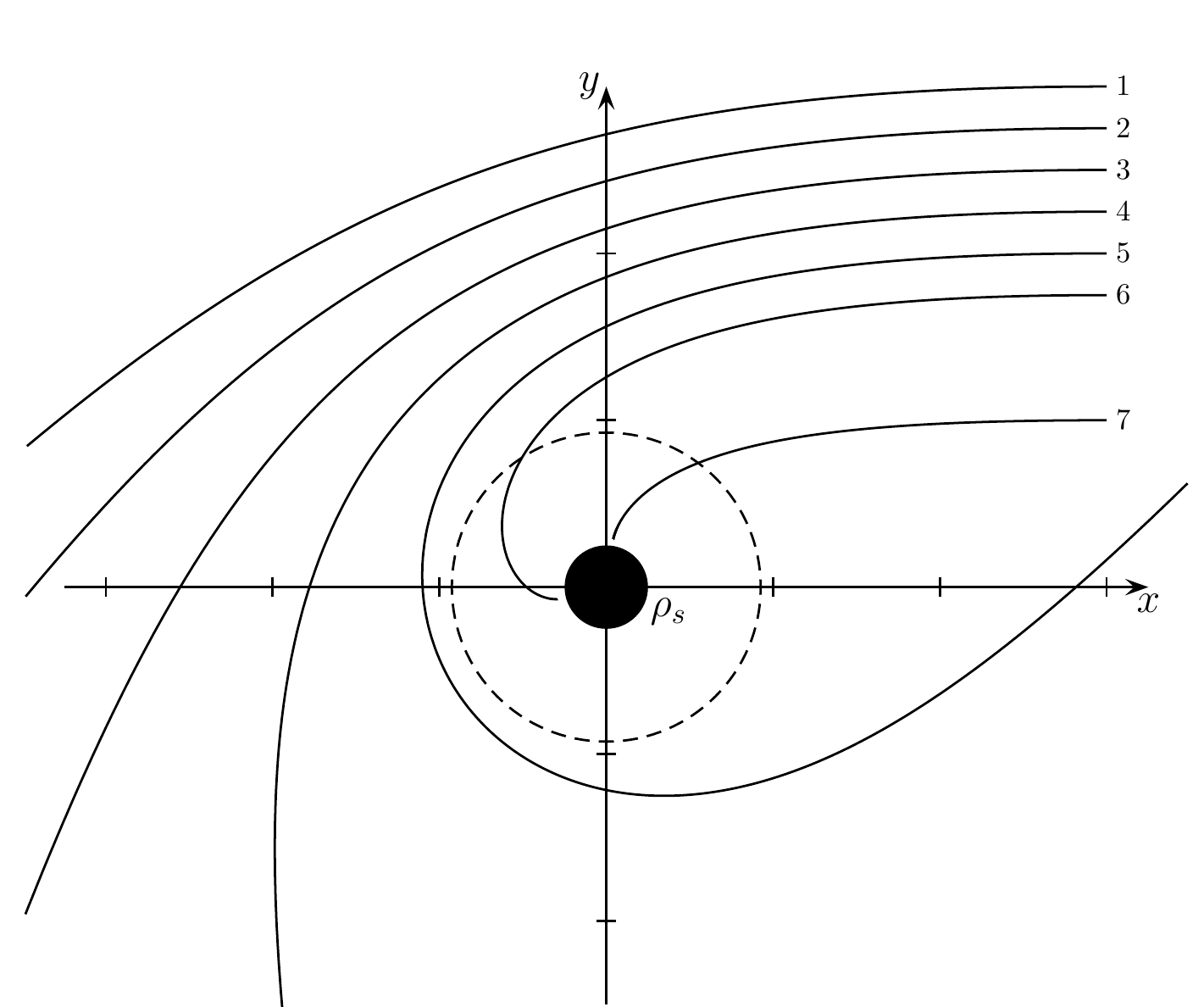}
 \caption{Light rays starting from $x=6$, $y_i=\left\{6,5.5,5,4.5,4,3.5,2\right\}$ in the negative $y$-direction are deflected by the curved space-time close to the Schwarzschild black hole (black disk). The dashed circle represents the photon orbit which represents the limit of the closest approach for light rays.}
 \label{fig:deflLightRays}
\end{figure}

\begin{gvResult}
 Figure~\ref{fig:deflLightRays} shows light rays for several initial positions $y_i$. The closer the light ray passes the black hole the stronger the rays are deflected. If the light ray crosses the photon orbit, $r_{\mbox{\tiny po}}=\frac{3}{2}r_s$ $(\rho_{\mbox{\tiny po}}=(2+\sqrt{3})\rho_s)$, the light ray cannot escape from the black hole. 
\end{gvResult}

\begin{gvConf}
 defl\_of\_light.
\end{gvConf}

\noindent A detailed explanation of the configuration file can be found in \ref{app:conf}.

It could also be interesting to find the initial position $y$ where the light ray approaches the photon orbit asymptotically. Here, we have $y\approx 3.9508265$. For that, however, it would be better to change the numerical integration from the standard Runge-Kutta fourth order integrator without step size control to the Runge-Kutta-Fehlberg integrator with step size control and absolute error tolerance of $\varepsilon_{\mbox{\tiny abs}}=10^{-12}$. Then, the constraint equation for the light ray is fulfilled until the light ray reaches the horizon.

\subsection{Einstein ring}
In the previous example, we have seen that for some specific initial positions $y$, the light ray orbits the black hole and returns to the observer.  Due to the symmetry of the Schwarzschild space-time, this happens also for light rays which lie in a plane that is rotated around the connecting axis between the observer and the black hole. As a result, the observer will see himself as an Einstein ring around the black hole. However, there is not only one but an arbitrary number of Einstein rings. 

One method to find the geodesics that return to the observer would be to solve the geodesic equations analytically, see e.g. M{\"u}ller~\cite{mueller2008prdB}. But this is extremely laborious even for the more adequate spherical coordinates. Another possibility, known from solving ordinary differential equations, is the shooting method which we can be easily simulated using the \GV{} and varying the initial direction $\xi$.

\begin{figure}[htb]
 \centering
 \includegraphics[scale=0.5]{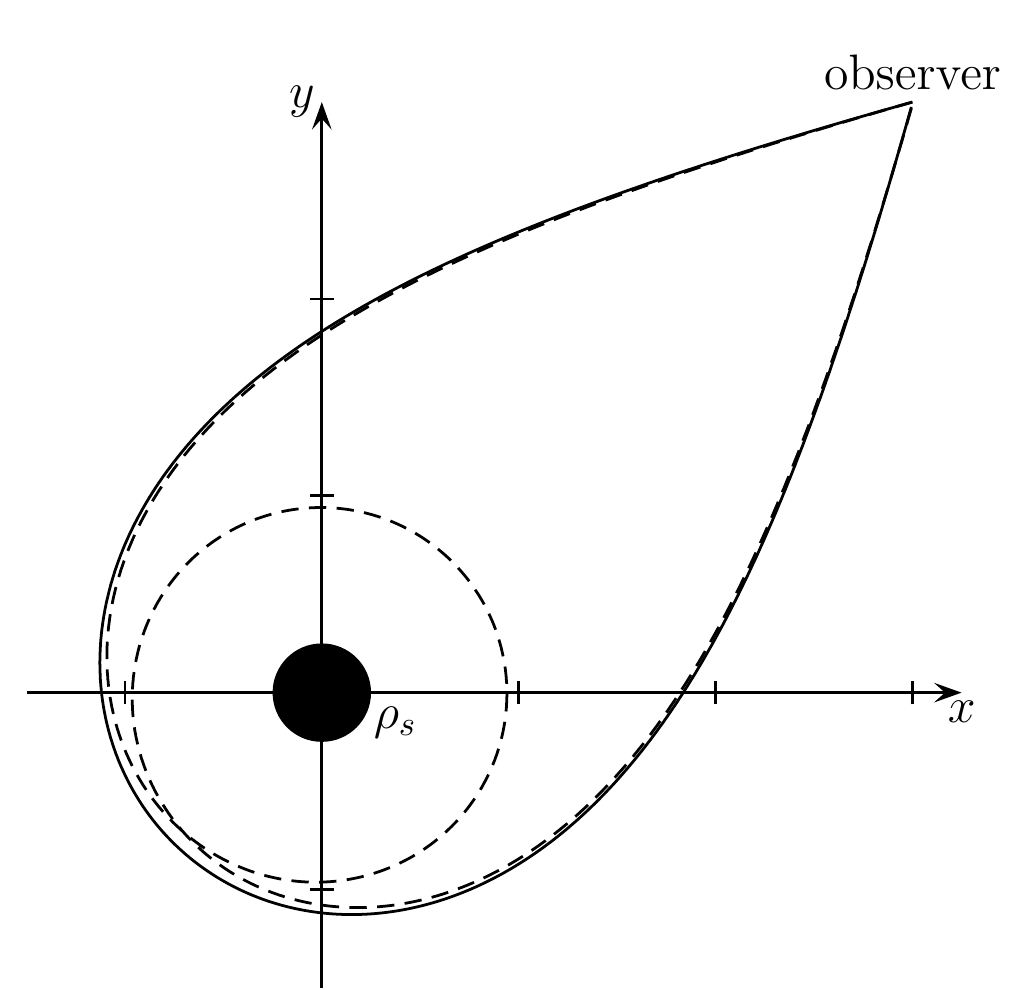}
 \caption{Light rays starting from the observer at $(x=6,y=6)$ return to him after orbiting the black hole once (solid line) or twice (dashed line). Hence, the observer will see himself as an Einstein ring of first or second order.}
 \label{fig:einsteinRings}
\end{figure}

\begin{gvExercise}
  Start the \GV{} and select the Schwarzschild metric in isotropic coordinates (SchwarzschildIsotropic) in the ``Metric/Integrator/Cons\-tants'' window. Set the position of an observer in the ``Local Tetrad'' window to $(t=0,x=6,y=6,z=0)$. Find the initial directions $\xi$ of the light rays that orbit the black hole once, twice, or thrice, cf. Fig.~\ref{fig:einsteinRings}, before returning to the observer. 
\end{gvExercise}

\begin{gvResult}
  The initial directions $\xi$ with respect to the observer's local reference frame, cf. Fig.~\ref{fig:initDir}, read: $\xi_1=195.668165\degree$, $\xi_2=195.96490458\degree$, and $\xi_3=195.965451625\degree$.
\end{gvResult}

\begin{gvConf}
  einstein\_ring\_1,einstein\_ring\_2,einstein\_ring\_3. 
\end{gvConf}

\begin{gvExercise}
  The situation is as before. But now, the observer sends out a flash of light and measures the time until the Einstein rings appear: $\tau_1\approx 40.1480$, $\tau_2\approx 69.2052$, $\tau_3\approx 98.2203$. What is the relation between proper time and coordinate time? Switch to the 2D view and use the coordinate time 't' as abscissa and 'x' as ordinate.
\end{gvExercise}

\begin{figure}[htb]
 \centering
 \includegraphics[scale=0.5]{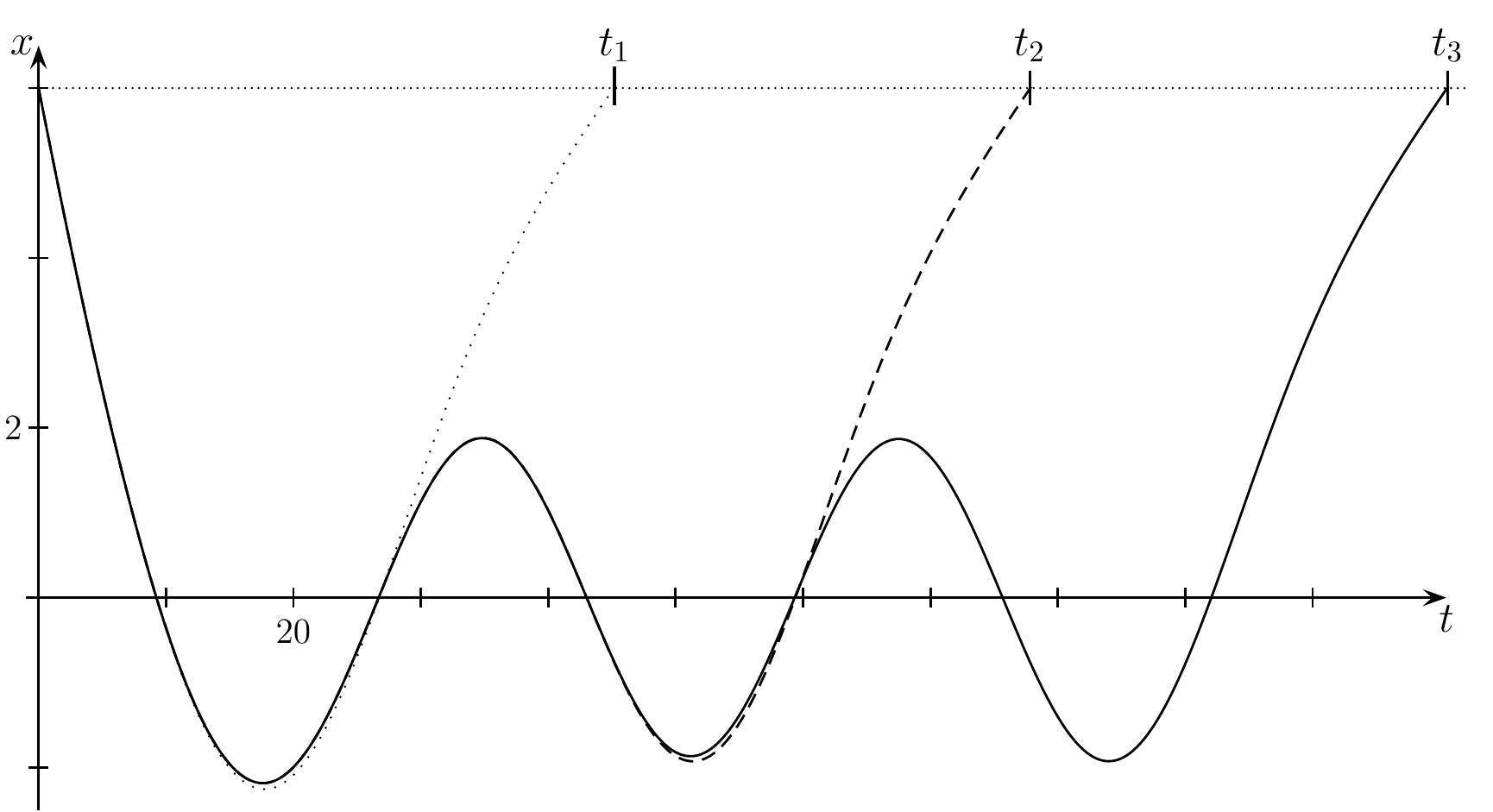}
 \caption{Light travel times for the light rays of Fig.~\ref{fig:einsteinRings} in 2D coordinate representation: $t_1\approx 45.1758$, $t_2\approx 77.8718$, $t_3\approx  110.5205$. }
 \label{fig:lightTravelTime}
\end{figure}

\begin{gvResult}
  From Fig.~\ref{fig:lightTravelTime} we can read the light travel times $t_i$ for the light rays with initial directions $\xi$. The relation between the coordinate time $t_i$ and the proper time $\tau_i$ for a fixed observer position is constant. 
\end{gvResult}

\begin{gvConf}
  light\_travel\_time.
\end{gvConf}

\noindent This example could be also formulated for given observer positions $\rho$ and corresponding light travel times $\tau(\rho)$ as measured by the observer. To prepare this exercise, the coordinate times $t$ for one orbit have to be determined in advance for several radial positions $\rho$. Then, by means of the relation $\tau/t=\sqrt{1-r_s/r}=\sqrt{1-4\rho_s/\left[\rho\left(1+\rho_s/\rho\right)^2\right]}$, which shall be found by the students, the proper times $\tau$ can be determined. Probably, this relation could be easier found in the standard spherical Schwarzschild coordinates.

\subsection{Shapiro time delay}
\begin{figure}[htb]
 \centering
 \includegraphics[scale=0.5]{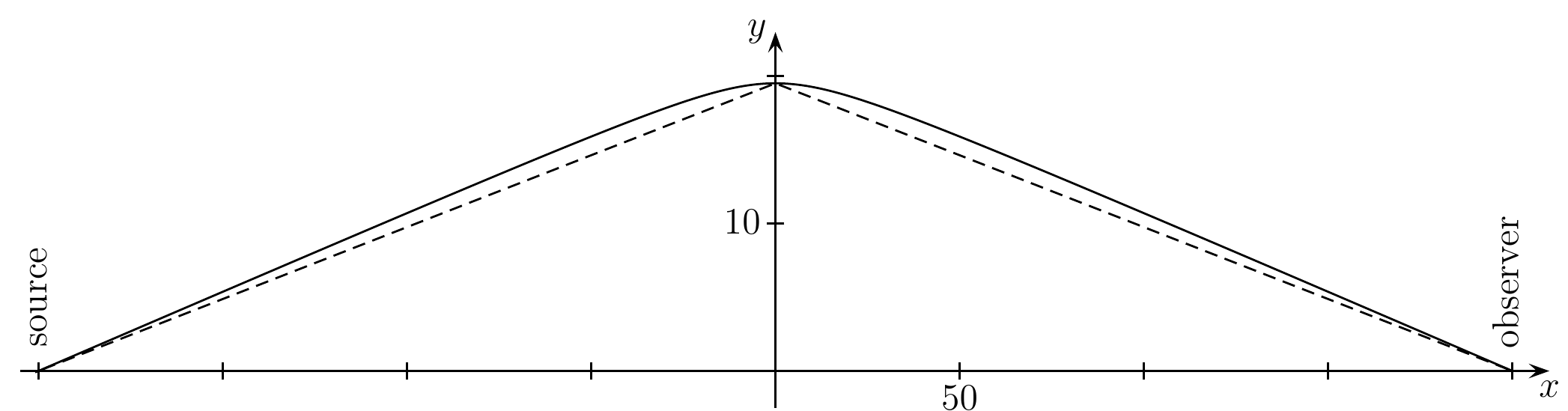}
 \caption{Shapiro time delay. The solid line represents the real light ray, whereas the dashed line is the rectilinear approximation. Impact parameter: $y=R$. Position of observer and source: $x_{\mbox{\tiny obs}}=X$, $x_{\mbox{\tiny source}}=-X$.}
 \label{fig:shapiro}
\end{figure}

In Ref.~\cite{rindler}, Rindler discusses the Shapiro time delay by means of a situation similar to the one shown in Fig.~\ref{fig:shapiro}. A light ray emitted at $x_{\mbox{\tiny source}}=-X$ passes a spherical mass $M$, which is located at $(x=0,y=0)$, at $y=R$ and reaches the observer at $x_{\mbox{\tiny obs}}=X$. From the isotropic metric, Eq.~(\ref{eq:schwMetricIso}), together with $ds=0$ and $\rho=\sqrt{X^2+R^2}$, he deduces the total coordinate time
\begin{equation}
 \Delta t\approx 2X+r_s\ln\frac{4X^2}{R^2},
  \label{eq:shapiro}
\end{equation}
where the logarithmic term is the Shapiro time delay.

\begin{gvExercise}
   Start the \GV{} and select the Schwarzschild metric in isotropic coordinates. Find a geodesic that connects the observer, located at $x{\mbox{\tiny obs}}=200$, with a light source at position $x_{\mbox{\tiny source}}=-200$. Compare the total light travel time with the approximation by Rindler. 
\end{gvExercise}

\begin{gvResult}
 The light ray that connects the observer and the light source in Fig.~\ref{fig:shapiro} has initial angle $\xi\approx 173.8578332\degree$. Rindler's approximation, Eq.~(\ref{eq:shapiro}), yields $\Delta t\approx 412.0797$, whereas from the \GV{} we obtain $\Delta t\approx 414.579$.
\end{gvResult}

\begin{gvConf}
 shapiro.
\end{gvConf}

\section{Periodic orbits of time-like geodesics in black hole space-times}\label{sec:kerrPeriodic}
The most simple periodic orbit of a time-like geodesic in the Schwarzschild space-time is a circular orbit. The relation between the radius $r$ of the orbit and the corresponding velocity $\beta$ can be easily found within the original, spherically symmetric Schwarzschild coordinates. In isotropic coordinates, this relation reads 
\begin{equation}
 \beta^{-1} = \sqrt{2\left[\frac{\rho}{4\rho_s}\left(1+\frac{\rho_s}{\rho}\right)^2-1\right]},
\end{equation}
which is valid only down to the last stable orbit, $\rho\geq\rho_{\mbox{\tiny lso}}=(5+2\sqrt{6})\rho_s$. However, beside these circular orbits, there are also more complex periodic orbits. Levin and Perez-Giz~\cite{levin2008} give a whole taxonomy of periodic orbits. See \ref{app:periodic} on how to reproduce the orbits shown in their paper.

In the following exercise, we will consider periodic orbits for Kerr black holes. In Boyer-Lindquist coordinates, see e.g. Bardeen et al.~\cite{bardeen1972}, the Kerr metric reads
\begin{eqnarray}
  \label{eq:kerrBLbardeen}
  ds^2 &= -\left(1-\frac{r_sr}{\Sigma}\right)c^2dt^2-\frac{2r_sar\sin^2\vartheta}{\Sigma}c\,dt\,d\varphi + \frac{\Sigma}{\Delta}dr^2 + \Sigma d\vartheta^2\\
      &\quad + \left(r^2+a^2+\frac{r_sa^2r\sin^2\vartheta}{\Sigma}\right)\sin^2\vartheta d\varphi^2,
\end{eqnarray}
with $\Sigma=r^2+a^2\cos^2\vartheta$, $\Delta=r^2-r_sr+a^2$, and $r_s=2GM/c^2$. $M$ is the mass and $a$ is the angular momentum per unit mass of the black hole. As the observer's local reference frame we use the locally non-rotating tetrad
\begin{eqnarray}
  \mathbf{e}_{(t)} &= \sqrt{\frac{A}{\Sigma\Delta}}\left(\frac{1}{c}\partial_t + \omega\partial_{\varphi}\right), \qquad& \mathbf{e}_{(r)} = \sqrt{\frac{\Delta}{\Sigma}}\partial_r,\\
  \mathbf{e}_{(\vartheta)} &= \frac{1}{\sqrt{\Sigma}}\partial_{\vartheta}, \qquad&  \mathbf{e}_{(\varphi)} = \sqrt{\frac{\Sigma}{A}}\frac{1}{\sin\vartheta}\partial_{\varphi},
\end{eqnarray}
where $\omega = r_sar/A$ and $A=\left(r^2+a^2\right)\Sigma+r_sa^2r\sin^2\vartheta$.

\begin{figure}[htb]
 \centering
 \includegraphics[scale=0.4]{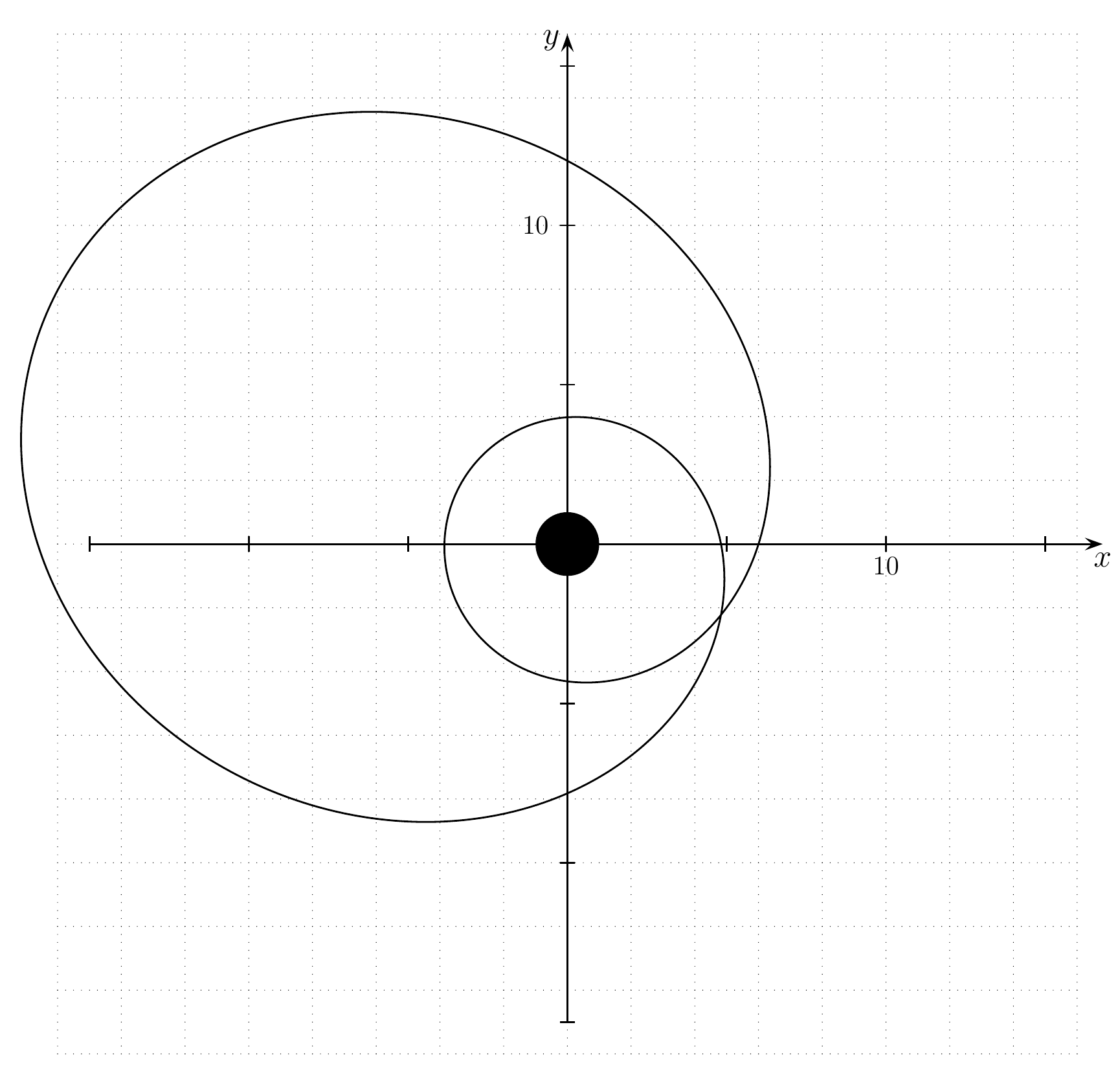}\includegraphics[scale=0.4]{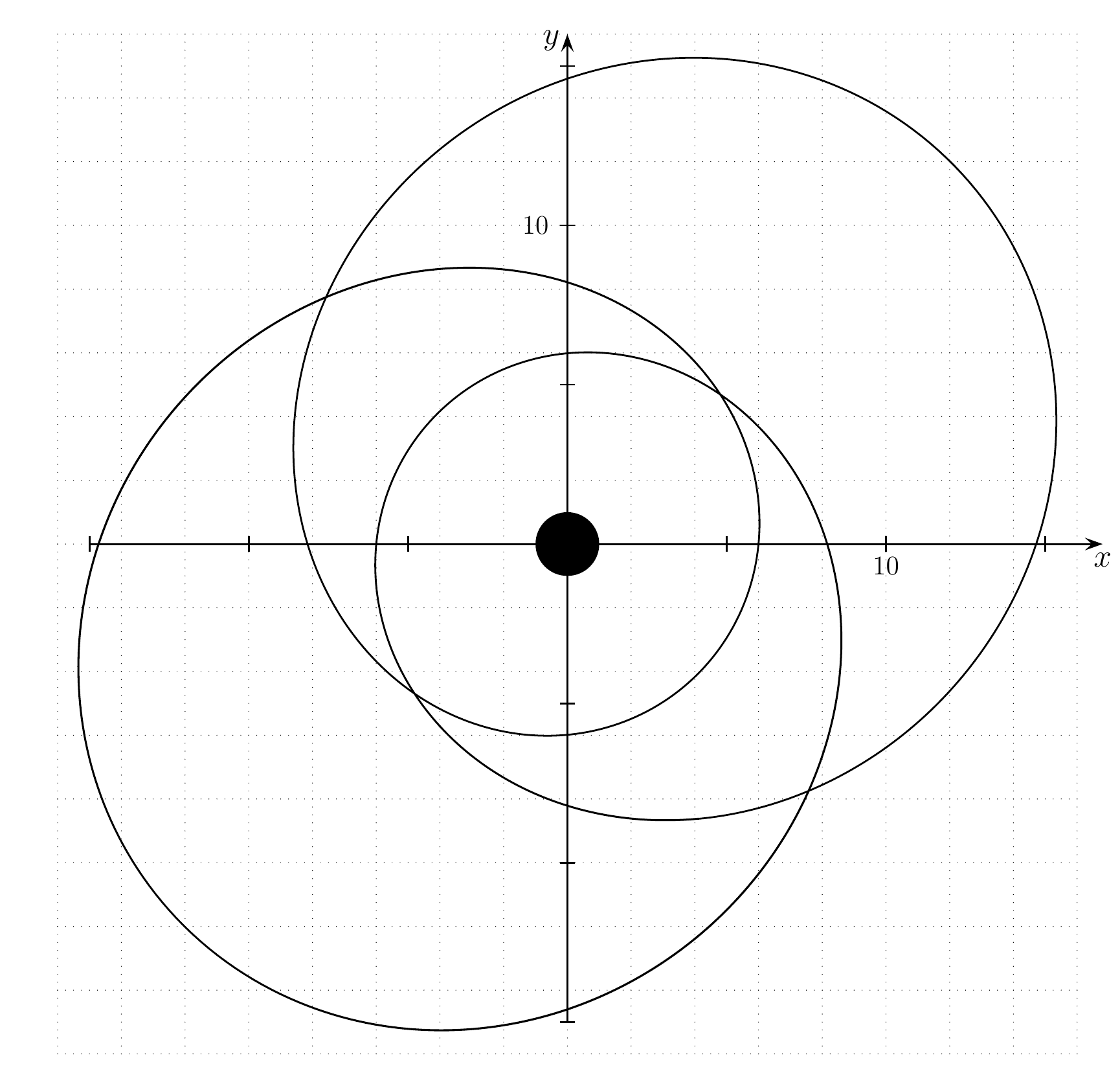}
 \caption{Periodic orbits in the Kerr space-time with parameters $M=1$ and $a=1/2$. The observer is located at $r=6,\varphi=0$ and starts a time-like geodesic with $\beta=1/2$ and $\chi=0\degree$, $\xi_1=67.65\degree$ (left), or $\xi_2\approx 82.95\degree$ (right).}
 \label{fig:kerrPeriodicOrbits}
\end{figure}

\begin{gvExercise}
  Select the Kerr metric in Boyer-Lindquist coordinates (KerrBL) and set the mass and angular momentum parameters to $M=1$, $a=1/2$. The observer is located at $(r=6,\vartheta=\pi/2,\varphi=0)$. With respect to his locally non-rotating reference frame, cf. 'Natural local tetrad' in the 'Local Tetrad' window, he starts a future directed time-like geodesic with initial velocity $\beta=1/2$ and direction $\chi=90\degree$, $\xi=67.649\degree$, see Fig.~\ref{fig:kerrPeriodicOrbits}. By varying the initial direction $\xi$, find periodic orbits of higher order. Instead of varying the initial direction, the initial velocity could also be changed.
\end{gvExercise}

\begin{gvResult}
 Periodic orbits of higher order follow from initial directions $\xi_2\approx 82.95\degree$, $\xi_3\approx 72.6\degree$, $\xi_4\approx 70.58\degree$, $\xi_5\approx 75.1\degree$, etc. For the fixed direction $\xi_1=67.65\degree$, we can also change the initial velocity: $\beta_2\approx 0.545$, $\beta_3\approx 0.52006$, $\beta_5\approx 0.528$.
\end{gvResult}

\begin{gvConf}
  kerr\_periodic.
\end{gvConf}

\section{Wormhole space-time}\label{sec:wormhole}
\begin{figure}[htb]
 \centering
 \includegraphics[scale=0.8]{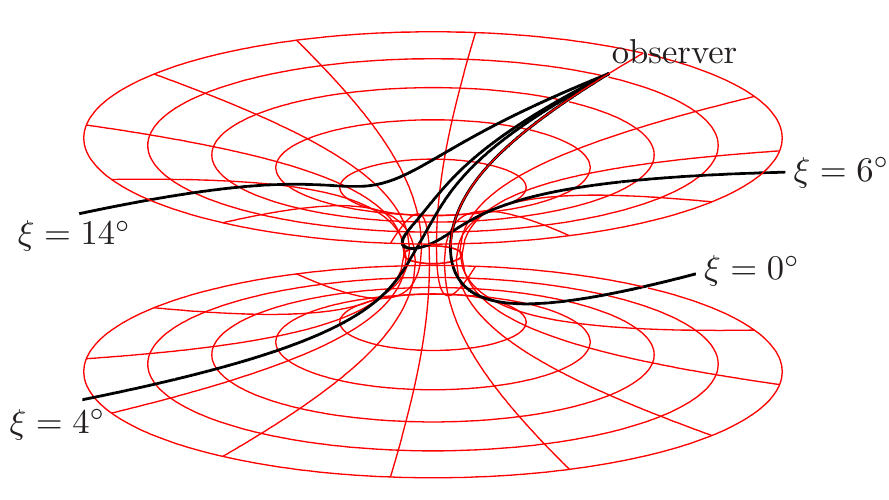}
 \caption{Embedding diagram of a Morris-Thorne wormhole space-time with $b_0=1$ and some exemplary null geodesics (black lines) with initial direction $\mathbf{k}=\mathbf{e}_{(t)}-\cos\xi\mathbf{e}_{(l)}+\cos\xi\mathbf{e}_{(\varphi)}$ starting at the observer's position $(l=10,\vartheta=\pi/2,\varphi=0)$.}
 \label{fig:mtEmbedding}
\end{figure}

The most simple non-trivial space-time is that of a Morris-Thorne~\cite{morris1988} wormhole whose metric, given in spherical coordinates, reads
\begin{equation}
  ds^2 = -c^2dt^2 + dl^2 + (b_0^2+l^2)\left(d\vartheta^2+\sin^2\!\vartheta\,d\varphi^2\right).
\end{equation}
Here, $b_0$ is the throat radius and $l$ is the proper radial coordinate. The local reference frame of a static observer is given by
\begin{equation}
 \fl\qquad \mathbf{e}_{(t)} = \frac{1}{c}\partial_t, \quad \mathbf{e}_{(l)} = \partial_l,\quad \mathbf{e}_{(\vartheta)} = \frac{1}{\sqrt{b_0^2+l^2}}\partial_{\vartheta}, \quad \mathbf{e}_{(\varphi)} = \frac{1}{\sqrt{b_0^2+l^2}\,\sin\vartheta}\partial_{\varphi}.
\end{equation}
To receive an impression of how the inner geometry of this wormhole space-time looks like, we take advantage of the spherical symmetry and embed the $(t=\mbox{const},\vartheta=\pi/2)$ hypersurface into the three-dimensional Euclidean space, cf. for example Ref.~\cite{mtw}. The resulting embedding function reads
\begin{equation}
  z(r) = \pm b_0\ln\left[\frac{r}{b_0}+\sqrt{\left(\frac{r}{b_0}\right)^2-1}\right],
\end{equation}
where the radial coordinates $r$ and $l$ are related via $r^2=b_0^2+l^2$. Figure \ref{fig:mtEmbedding} shows an embedding diagram for a wormhole with throat size $b_0=1$ and an observer at $(l=10,\vartheta=\pi/2,\varphi=0)$.

\begin{gvExercise}
 Select the Morris-Thorne wormhole and set the throat size parameter $b_0=1$. The observer is located at $(l=10,\vartheta=\pi/2,\varphi=0)$ and uses the inward-oriented tetrad $\mathbf{e}_{(0)}=\mathbf{e}_{(t)}$, $\mathbf{e}_{(1)}=-\mathbf{e}_{(l)}$, $\mathbf{e}_{(2)}=\mathbf{e}_{(\varphi)}$, $\mathbf{e}_{(3)}=\mathbf{e}_{(\vartheta)}$ as local reference frame. A light ray with initial direction $\chi=90\degree, \xi_0=0\degree$ traverses the wormhole and hits the point $P=(l=-10, \varphi=0)$. Now, find the initial direction $\xi_1$ where the new light ray hits the point $P$ again but now with $\varphi=2\pi$. Explain what an observer would see between $\xi_0$ and $\xi_1$.
\end{gvExercise}

\begin{gvResult}
 A light ray with initial angle $\xi\approx 5.64336\degree$ travels around the wormhole throat before reaching the point $P$. Between $\xi_0$ and $\xi_1$ the observer would see the whole lower universe. Here, an interactive tool is indispensable to comprehend this fact.
\end{gvResult}

\begin{gvConf}
 morristhorne.
\end{gvConf}

\noindent Detailed discussions for the first-person visualizations of the Morris-Thorne wormhole can be found in M{\"u}ller~\cite{mueller2004} or Ruder~\cite{ruder2008}.

\section{Multi black hole solution}\label{sec:multiBH}

Interesting general relativistic situations occur when several black holes are combined. These configurations will be inherently dynamical due to the mutual gravitational  attraction between the black holes. However, it is possible to construct static situations if one allows the black holes to carry an electric charge. Then the gravitational attraction can be compensated exactly by the electric repulsion. These solutions have been discovered by Majumdar~\cite{majumdar1947} and Papapetrou~\cite{papapetrou1947}. They can be described as a collection of any number of extreme Reissner-Nordstr{\o}m black holes. We consider here the case of only two
such black holes, see Chandrasekhar~\cite{chandrasekhar1989a},

In Cartesian coordinates, the extreme Reissner-Nordstr{\o}m metric reads
\begin{equation}
 ds^2 = -\frac{dt^2}{U^2}+U^2\left(dx^2+dy^2+dz^2\right),
\end{equation}
where $U=1+M_1/r_1+M_2/r_2$, $r_1=\sqrt{x^2+y^2+(z-1)^2}$, and $r_2=\sqrt{x^2+y^2+(z+1)^2}$. Here, we use geometric units with $G=c=1$. The two black holes are located at $x=y=0, z=\pm 1$. As local reference frame, we use
\begin{eqnarray}
 \mathbf{e}_{(t)}=U\partial_t,\quad \mathbf{e}_{(x)}=\frac{1}{U}\partial_x, \quad \mathbf{e}_{(y)}=\frac{1}{U}\partial_y, \quad \mathbf{e}_{(z)}=\frac{1}{U}\partial_z.
\end{eqnarray}
For simplicity, we set $M_1=M_2=1$.
\begin{figure}[htb]
 \centering
 \includegraphics[scale=1.0]{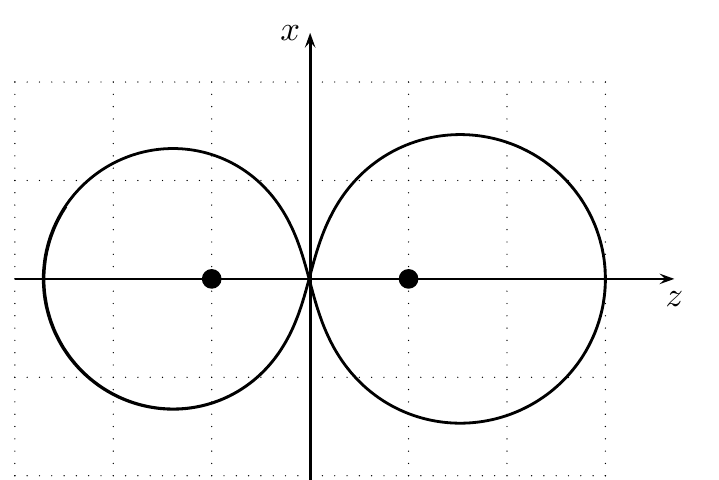}\,\includegraphics[scale=1.0]{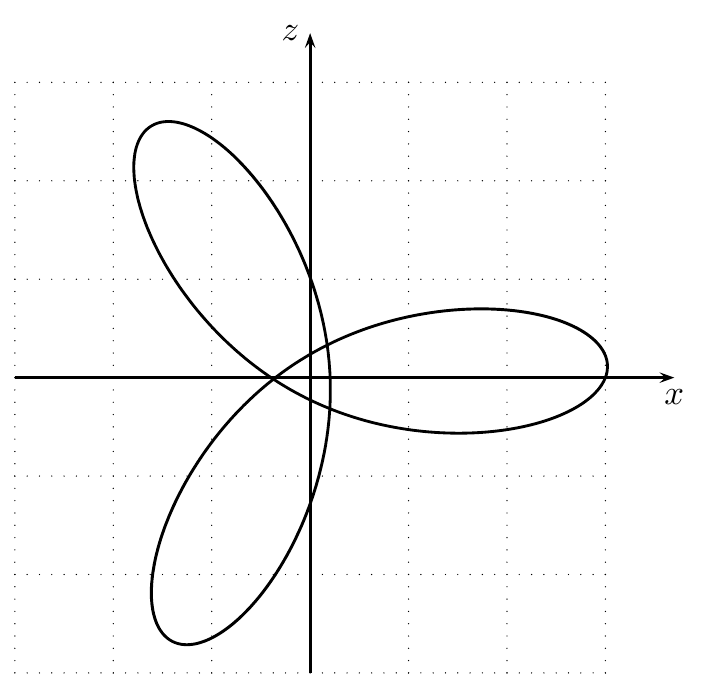}
 \caption{Time-like periodic orbits in the extreme Reissner-Nordstr{\o}m metric with the two black holes located at $z=\pm 1$. \textbf{Left:} The geodesic starts at $(x=y=0,z=3)$ with four-velocity $\mathbf{u}=\gamma\left(\mathbf{e}_{(t)}+\beta\mathbf{e}_{(x)}\right)$, where $\beta\approx 0.591943$ and $\gamma=1/\sqrt{1-\beta^2}$. \textbf{Right:} The geodesic starts at $(y=z=0,x=3)$ with four-velocity $\mathbf{u}=\gamma\left(\mathbf{e}_{(t)}+\beta\cos\xi\mathbf{e}_{(x)}+\beta\sin\xi\mathbf{e}_{(y)}\right)$, where $\beta=0.2$ and $\xi\approx 69.26$.}
 \label{fig:extrRN}
\end{figure}

\begin{gvExercise}
 Select the extreme Reissner-Nordstr{\o}m metric (ExtremeReissnerNordstromDihole) and set the mass parameters of the two black holes to $M_1=M_2=1$. Find time-like periodic orbits in the $xy$- and $xz$-planes.
\end{gvExercise}

\begin{gvResult}
 Some exemplary time-like orbits are the following: \textbf{1)} $(x=0,y=0,z=3)$, $\mathbf{u}=\gamma\left(\mathbf{e}_{(t)}+\beta\mathbf{e}_{(x)}\right)$, $\beta\approx 0.896899$; \textbf{2)} $(x=0,y=0,z=3)$, $\mathbf{u}=\gamma\left(\mathbf{e}_{(t)}+\beta\mathbf{e}_{(x)}\right)$, $\beta\approx 0.591943$; \textbf{3)} $(x=0,y=0,z=10)$, $\mathbf{u}=\gamma\left(\mathbf{e}_{(t)}+\beta\mathbf{e}_{(x)}\right)$, $\beta\approx 0.451$; 

 A light-like orbit is given by $(x=0,y=0,z\approx 1.720218)$, $\mathbf{k}=\mathbf{e}_{(t)}+\mathbf{e}_{(x)}$. 
\end{gvResult}

\begin{gvConf}
 extrRNdihole1, extrRNdihole2, extrRNdihole3, extrRNdihole4, extrRNdihole5.
\end{gvConf}

\section{Summary and outlook}
The \GV{} is a valuable tool to study the behaviour of light-like and time-like geodesics in space-times whose metrics are provided analytically. While geodesics could also be visualized using standard software, a clear understanding of their behaviour can only be achieved by interactively manipulating the corresponding parameters. 

In this paper, we have given a small number of possible applications formulated as exercises for students in the classroom. This can only hint at various other possibilities. The strength of \GV\ lies in its visualization capabilities and to a large part in the available database of exact solutions. This allows the user to study the behaviour of geodesics in many different circumstances. The possibility to choose the curve parameters in an interactive way means that one can make numerous parameter studies depending on the problem under consideration.

So far, the \GV{} can only handle one geodesic at once. This limitation will be overcome in a future version. We also plan to make the \GV{} scriptable to realize more complex demonstrations.

\appendix
\section{Configuration file example for the GeodesicViewer}\label{app:conf}
As an example, we give here the configuration file for the 'Deflection of light' exercise of Sec.~\ref{subsec:deflection} as written by the \GV{}. Although this file is a plain text file, we do not recommend to modify it by hand. Besides, note that not all of these parameters are crucial for this exercise. 
{\small
\begin{verbatim}
   --------------------------------------------------------------------
   METRIC       SchwarzschildIsotropic
   PARAM  0        mass    1.000000000000
   INIT_POS           0.00000000   6.00000000   6.00000000   0.00000000
   INIT_DIR          -1.00000000   0.00000000   0.00000000
   INIT_ANGLE_VEL   180.00000000  90.00000000   0.99000000
   TIME_DIR          1
   AXES_ORIENT       0
   GEOD_SOLVER_TYPE  4
   GEODESIC_TYPE     lightlike
   STEPSIZE_CTRL     1
   STEPSIZE          1.00000000e-02
   STEPSIZE_MAX      1.00000000e+00 
   EPSILONS          1.00000000e-12 0.00000000e+00
   CONSTR_EPSILON    1.00000000e-06
   MAX_NUM_POINTS    3000
   TETRAD_TYPE       0
   BASE_0            1.00000000   0.00000000   0.00000000   0.00000000
   BASE_1            0.00000000   1.00000000   0.00000000   0.00000000
   BASE_2            0.00000000   0.00000000   1.00000000   0.00000000
   BASE_3            0.00000000   0.00000000   0.00000000   1.00000000
   BOOST             0.00000000  90.00000000   0.00000000
   SPEED_OF_LIGHT        1.000000
   GRAV_CONSTANT     1.000000e+00
   DIELECTRIC_PERM   1.000000e+00
   --------------------------------------------------------------------
\end{verbatim}
}
\noindent Most of the parameters are self-explanatory. The time-direction (\verb+TIME_DIR+) can be either plus or minus one depending on the geodesic being future- or past-directed. The method for solving the geodesic equation (\verb+GEOD_SOLVER_TYPE+) is encoded by just a number, cf. \verb+m4dMotionList.h+ of the Motion4D library. The absolute and relative error tolerances are given by the \verb+EPSILONS+ values. The numerical integration of the geodesic stops when the constraint equation with \verb+CONSTR_EPSILON+ is no longer fulfilled. For the local reference frame in this exercise, we use the default natural local tetrad. Hence, the \verb+TETRAD_TYPE+ is set to zero, the base vectors \verb+BASE_0+ to \verb+BASE_3+ build the identity matrix, and there is no boost transformation (the last value is the boost velocity). Since we use geometrical units, the speed of light and the gravitational constant are set to unity.

\section{Periodic orbits by Levin and Perez-Giz}\label{app:periodic}
The periodic orbits in the Schwarzschild space-time shown in the paper by Levin and Perez-Giz~\cite{levin2008}, can be reproduced in isotropic coordinates in the following way. For a nearly arbitrary radial coordinate $r$, the initial velocity $\beta$ and the initial direction $\xi$ read
\begin{equation}
 \beta = \sqrt{1-\frac{1-r_s/r}{E^2}},\quad \xi=\arcsin\frac{L}{r\sqrt{E^2/(1-r_s/r)-1}}.
\end{equation}
The isotropic position is given by
\begin{equation}
 x = \frac{1}{4}\left(2r-r_s+2\sqrt{r(r-r_s)}\right).
\end{equation}
For example, to reproduce the periodic orbit of Fig.~12 in their paper with parameters $L=3.9$ and $E=0.987160$, we set $r=10$ and obtain $\beta\approx 0.423147$, $\xi\approx 56.62467$, and $x\approx 8.972136$.


\ack
This work has been partly supported by the Marsden Fund of the Royal Society of New Zealand under contract number UOO0922.

\section*{References}
\bibliographystyle{unsrt}
\bibliography{lit_geod}

\end{document}